\documentclass[9pt,twocolumn,twoside,lineno]{pnas-new}
\templatetype{pnasresearcharticle} % Choose template 
\title{Female scholars need to achieve more for equal public recognition}

\author[a]{Menno H. Schellekens}
\author[b]{Floris Holstege}
\author[a,c,1]{Taha Yasseri} 

\affil[a]{Oxford Internet Institute, University of Oxford}
\affil[b]{Leiden University College}
\affil[c]{Alan Turing Institute for Data Science and Artificial Intelligence}

\leadauthor{Schellekens} 

\correspondingauthor{\textsuperscript{1}To whom correspondence should be addressed. E-mail: taha.yasseri@oii.ox.ac.uk}


\begin{abstract}
\nolinenumbers
 Different kinds of "gender gap" have been reported in different walks of the scientific life, almost always favouring male scientists over females. In this work, for the first time, we present a large-scale empirical analysis to ask whether female scientists with the same level of scientific accomplishment are as likely as males to be recognised. We particularly focus on Wikipedia, the open online encyclopedia that its open nature allows us to have a proxy of community recognition. We calculate the probability of appearing on Wikipedia as a scientist for both male and female scholars in three different fields. We find that women in Physics, Economics and Philosophy are considerable less likely than men to be recognised on Wikipedia across all levels of achievement.
\end{abstract}

\dates{This manuscript was compiled on \today}
\begin{document}
\maketitle
\thispagestyle{firststyle}
\ifthenelse{\boolean{shortarticle}}{\ifthenelse{\boolean{singlecolumn}}{\abscontentformatted}{\abscontent}}{}
\nolinenumbers
Female scholars face many more barriers in their professional path compared to their male colleagues \citep{Raymond2013,Lariviere2013}. They have been found to be discriminated against in the workplace \citep{Clancy2014}, in grant applications \citep{Bornmann2007, Boyle2015}, and as students \citep{Nosek2009,Pollack2013,Yang2033}, and they experience sexism on daily basis \cite{melville2019}. Apart from systematic biases and barriers against female scholars, that might be a reflection of a wider societal issue, the community of academics themselves might be suffering from prejudice and negative perceptions on female scholars. The question that we ask in this work is if female scholars are less likely to be recognised by their communities for their accomplishments. 

Wikipedia, the largest crowd-based knowledge repository has been studied from different angles. There are arguments about its accuracy \citep{giles2005internet}, coverage \citep{halavais2008analysis}, and neutrality \citep{greenstein2012wikipedia}, in both directions. 
Previous work has reported that the level of attention given to scholars measured by the level of activity and the traffic to the Wikipedia articles that are dedicated to them is not in proportion to their scholarly achievements evaluated by scientometrics measures \citep{samoilenko2014distorted}. Wikipedia traffic has been used as a proxy for collective attention \cite{garcia2016} and collective memory \cite{garcia2017memory}. However, in this work we use Wikipedia as a proxy for community recognition of academics and simply measure the difference of chances of being featured on Wikipedia for males and females who have similar scientific achievements. We build on previous work that generally reported that Wikipedia suffers from the lack of entries about female scientists \citep{Devlin2018, Wade2018} and lower quality and information reach of articles about women in general \citep{reagle2011gender,halfaker2017interpolating,graells2015first}. However, in addition to the existing case reports, here, we take a systematic approach by analysing the data on 15,049 scholars from three different disciplines. See Table~\ref{tab:sumstatscat} for details.

\begin{table}[ht]
\centering
  \caption{Overview of the scholars in the dataset} 
  \label{tab:sumstatscat} 
\begin{tabular}{lllrrr}
  %\hline
 Field & Gender & Count & \% on Wikipedia \\%& \% not on Wikipedia \\ 
  \hline
Physics  & female & 642 & 8.26 \\%& 91.74 \\ 
 & male & 5448 & 14.12 \\%& 85.88 \\ 
Economics & female & 1586 & 8.32 \\%& 91.68 \\ 
 & male & 5477 & 17.89 \\%& 82.11 \\ 
Philosophy & female & 467 & 15.42 \\%& 84.58 \\ 
  & male & 1429 & 28.06 \\%& 71.94 \\ 
   \hline
Total & female & 2695 & 9.54 \\%& 90.5 \\ 
 & male & 12354 &17.4 \\%& 82.6\\ 
   \hline  \hline
   Grand Total & & 15049 & 15.99 \\%& 84.01\\
 \end{tabular}
\end{table}
The fundamental question that we are addressing here is if there are fewer Wikipedia entries about female scientists \citep{reagle2011gender} because few women enter the sciences, or because they are less likely to contribute groundbreaking research, or do they face additional hurdles in attaining public recognition for their work for the same level of achievement? To answer this question, we analyse a dataset that is collected from Google Scholar and check it against Wikipedia entries. This dataset allows us to compare whether gender influences the chance of having a dedicated Wikipedia entry, controlling for scientific achievement. We find strong evidence of  discrimination in public recognition of scientific achievement gauged by  inclusion in Wikipedia at any level of success. %Specifically, women are between 17\% and 40\% less likely to receive a Wikipedia entry than male peers when both have an average h-index of their field, depending on the field.

While barriers for women in science at different stages of their careers have been reported and discussed intensively, there is little empirical work on the recognition of scientific achievement by the general public. This paper contributes the first large scale empirical analysis of gender bias in recognition of scientific accomplishments. 
 
\section*{Results\label{sec:results}}
We employ logistic regression to test the relationship between gender and Wikipedia recognition, controlling for h-index. For details of data collection and gender detection see Materials and Methods. We start with a simple model that has the following structure
\begin{equation}
  p(W) = B_1g + B_2h,
\end{equation}
where $W$ denotes the existence of a Wikipedia page, $g$ denotes the gender of the scientist, $h$ their h-index, and $B_1$ and $B_2$ are the model constants. This model assumes that being male or female changes the chance of recognition irrespective of academic achievement.  
In the nest step, considering that scientific accomplishments by females might be viewed differently, we add a new term to the model with an interaction between h-index and gender, 
\begin{equation}
p(W) = B_1g + B_2h + B_3 gh.    
\end{equation}

% Table created by stargazer v.5.2.2 by Marek Hlavac, Harvard University. E-mail: hlavac at fas.harvard.edu
% Date and time: Fri, Apr 12, 2019 - 11:25:23
\begin{table}[!htbp] \centering 
  \caption{} 
  \label{tab:regress} 
\begin{tabular}{@{\extracolsep{5pt}}lccc} 
\\[-1.8ex]\hline 
\hline \\[-1.8ex] 
 & \multicolumn{3}{c}{\textit{Dependent variable: Wikipedia page exists}} \\ 
%\cline{2-4} 
%\\[-1.8ex] & \multicolumn{3}{c}{Wikipedia page exists} \\ 
\\[-1.8ex] & (1) & (2) & (3)\\ 
\hline \\[-1.8ex] 
 Male & 0.480$^{***}$ & 0.557$^{***}$ & 2.139$^{***}$ \\ 
  & (0.071) & (0.073) & (0.313) \\ 
  & & & \\ 
 Logged h-index & 0.581$^{***}$ & 1.022$^{***}$ & 1.502$^{***}$ \\ 
  & (0.032) & (0.037) & (0.098) \\ 
  & & & \\ 
 Field: Economics &  & 0.807$^{***}$ & 0.800$^{***}$ \\ 
  &  & (0.055) & (0.055) \\ 
  & & & \\ 
 Field: Philosophy &  & 2.053$^{***}$ & 2.056$^{***}$ \\ 
  &  & (0.081) & (0.081) \\ 
  & & & \\ 
 Male:Logged h-index &  &  & $-$0.541$^{***}$ \\ 
  &  &  & (0.102) \\ 
  & & & \\ 
 Constant & $-$3.829$^{***}$ & $-$5.905$^{***}$ & $-$7.292$^{***}$ \\ 
  & (0.112) & (0.147) & (0.310) \\ 
  & & & \\ 
\hline \\[-1.8ex] 
Observations & 15,049 & 15,049 & 15,049 \\ 
Log Likelihood & $-$6,385.949 & $-$6,063.457 & $-$6,048.604 \\ 
Akaike Inf. Crit. & 12,777.900 & 12,136.920 & 12,109.210 \\ 
\hline 
\hline \\[-1.8ex] 
\textit{Note:}  & \multicolumn{3}{r}{$^{*}$p$<$0.1; $^{**}$p$<$0.05; $^{***}$p$<$0.01} \\ 
\end{tabular} 
\end{table} 

The results from the fit of the model to the data presented in Table~\ref{tab:regress} point towards structural discrimination in the recognition of scientific achievement. Regardless of field of study, being male significantly increases the chance of being recognised and featured on Wikipedia.

The negative interaction effect between gender and h-index suggests that Wikipedia's bias towards men is strongest amongst scientists with relatively low indexes. Gender plays a smaller role in the recognition of academics with exceptional academic standing.

\begin{figure*}
\centering
\includegraphics[width=.9\linewidth]{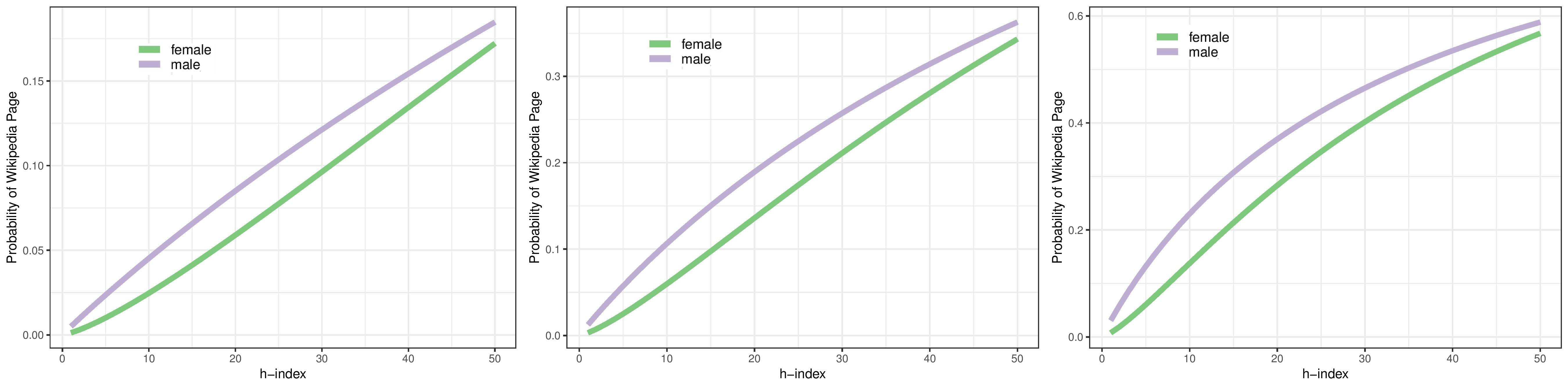}
\caption{Probability of male (purple) and female (green) scholars getting a Wikipedia page at different levels of scientific standing for Physics, Economics, and Philosophy, from left to right. Error bars are too small to be visible.}\label{fig:frog}
\end{figure*}

Logistic regression produces log-odds as coefficients; using those, we have plotted the probability that an economist of both genders is recognised with a Wikipedia page at different h-index levels (Figure~\ref{fig:frog}). A female economist with an average h-index has a probability of 0.11 of being recognised by Wikipedia, while an average male economist has a probability of 0.18. A male economist has to achieve an h-index of 11 for a similar probability of public recognition as a female economist with an h-index of 19. Similar patterns are observed for Physics and Philosophy.
Women are 19\%, 37\% and 50\% less likely to receive recognition than male peers when both have an average h-index in Physics, Economics and Philosophy respectively. We calculate these percentages by dividing the predicted probability of a women with an average h-index of having a Wikipedia page by the predicted probability for a man with the same h-index to have a Wikipedia page in the same field of research. 

To check the robustness, we provide a number of variations of this model to see if the effects hold. To control for cross-discipline differences, we run separate models per field to investigate the differences between fields (see Table~ S1).  
To further check the robustness of the results, we use alternative measures for scientific achievement such as  raw number of citations and h5 index to test if that changes the outcomes (see Table~S2).
This finding holds when controlling for field (see Table \ref{tab:regress}), when run separately for every field (see Table~S1) and when alternative measures are used (see Table~S2). It is statistically significant at $p < 0.01$ in all analyses.

\section*{Discussions}
We report on evidence of a bias against recognising the scientific accomplishments of women on Wikipedia. Men are more likely to be awarded a page in the world's most influential encyclopedia than women with similar scientometric records. This finding is replicated in Physics (natural sciences), Economics (social sciences), and Philosophy (humanities). The magnitude of male advantage is remarkably similar across the disparate fields.

It is beyond the scope of this paper to establish the causal mechanism behind the gender gap in recognition. Is research from women taken less seriously? Are males more easily given access to public fora to discuss their findings? And one should note that the biases reported in this work are on top of the reported biases on research funding allocations \cite{head2013differences}, publishing practices and hiring exercises \cite{carr2003ton,clauset2015systematic}.

We must also note that a portion of the reported bias might be due to the known gender gap among Wikipedia editors. It is notable that there are few female editors amongst the ranks of Wikipedia editors \citep{Wade2018,hill2013wikipedia}. The Wikimedia Foundation might want to consider policy changes to give women equal recognition for equal work as an starting point to battle this societal malfunction in a wider scope.

\section*{Materials and Methods\label{sec:data}}
The analysis is conducted with three measures: scientific accomplishment (retrieved with Google Scholar), gender (retrieved from \url{genderize.io}) and recognition from Wikipedia (retrieved from the Wikipedia API). We will cover each measure in the following sections. The summary statistics are available in Table~\ref{tab:sumstatscat} and in Table~S3.

\subsection{Scientific Accomplishment}
While scientists receive many forms of recognition, the most common measure is the citation. Citation metrics have increased in importance in the scientific realm. The h-index is widely preferred over raw citation counts, because it accounts for both the number of publications and citations \citep{kelly2006h}.
Many universities set minimum h-index values for new hires, and some universities base promotions on h-index thresholds \citep{Hicks2015}. 

The source of our dataset is Google Scholar. We queried a particular field and collected names in the order Google Scholar presents them, which is ordered by citation count. For every scientist, we retrieved citation counts, their name and their institution. Google Scholar has been found to have the largest coverage as compared to other databases, with up to 33\% more authors than its direct competitors and more diverse publications, such as conference papers and books \citep{Meho2007,Kousha2011, Harzing2013}. Thus, we are satisfied that Google Scholar gives an accurate and comprehensive overview of active scholars and their citations. 

Collecting data from Google Scholar is laborious, so we sampled scholars from three fields in different parts of the academic world: Physics (natural sciences), Economics (social sciences) and Philosophy (humanities) \citep{Holstege2018}. As reported in Table \ref{tab:sumstatscat}, the number of scholars in our sample, the gender balance and proportion of scholars with Wikipedia pages differs per field. 

Our sample of scholars is non-random, because scientists were ordered by h-index. However, we collected the top 10,000 available scholars from a field. The `bottom' of our sample contains scholars with h-indices as low as 1, so we cover a wide range of achievement. If we missed scholars, they must have very few citations and publications. This does not compromise our analysis, because these scholars are not likely to receive recognition from Wikipedia and thus not relevant. All three citation measures are not normally distributed (See Figures S1-S3) and transformed for the regression analysis.

\subsection{Gender}
Google Scholar does not list the gender of a scientist. Therefore, we must detect the gender of a scholar based on their first name. This technique is widely used and accurate \citep{Lariviere2013, Karimi2016}. We use genderize.io API, which makes use of a database of 216286 names from 79 countries and 89 languages to make prediction. Conveniently, genderize.io reports the number of times a given name appears in their database and the proportion of the two sexes. We applied strict filters: only predictions with a confidence greater than 90\% based on a minimum sample size of 10 were accepted. This measure cut our sample down to 15,049 from 23.000 scholars collected via Google Scholar. 

Genderize makes the assumption that persons who are a woman identify as female. However, both sexes can identify as many genders. The analysis would be superior if we could use the identified genders of every scientist, but this possibility is  not available. Given that it is common for women to identify as female and men as male, we use the Genderize categorization as the closest available proxy.

\subsection{Recognition by Wikipedia}
We queried the Wikipedia API with the names of scholars to check for Wikipedia pages under their name. When the Wikipedia page is listed under a slightly different name or a known alias, the Wikipedia API automatically refers us to the correct page. We checked a sample of 30 codings manually and found no miscodings.

\acknow{%Please include your acknowledgments here, set in a single paragraph. Please do not include any acknowledgments in the Supporting Information, or anywhere else in the manuscript.
We thank Jop Flameling for discussion on the research design and data collection. TY was partially supported by the Alan Turing Institute under the EPSRC grant no. EP/N510129/1.}

\showacknow{} % Display the acknowledgments section

% Bibliography
\bibliography{inequality}
\onecolumn
\section*{Supplementary Information}
\setcounter{figure}{0}
\makeatletter 
\renewcommand{\thefigure}{S\@arabic\c@figure}
\makeatother

\setcounter{table}{0}
\makeatletter 
\renewcommand{\thetable}{S\@arabic\c@table}
\makeatother

\begin{table}[!htbp] \centering 
  \caption{Regression Results for (1) Physics, (2) Economics and (3) Philosophy} 
  \label{tab:byfield} 
\begin{tabular}{@{\extracolsep{5pt}}lccc} 
\\[-1.8ex]\hline 
\hline \\[-1.8ex] 
 & \multicolumn{3}{c}{\textit{Dependent variable:}} \\ 
\cline{2-4} 
\\[-1.8ex] & \multicolumn{3}{c}{wiki\_bool} \\ 
\\[-1.8ex] & (1) & (2) & (3)\\ 
\hline \\[-1.8ex] 
 gendermale & 0.493$^{***}$ & 0.588$^{***}$ & 0.474$^{***}$ \\ 
  & (0.150) & (0.101) & (0.148) \\ 
  & & & \\ 
 log(h.index) & 0.746$^{***}$ & 1.236$^{***}$ & 1.011$^{***}$ \\ 
  & (0.064) & (0.057) & (0.076) \\ 
  & & & \\ 
 Constant & $-$4.867$^{***}$ & $-$5.773$^{***}$ & $-$3.757$^{***}$ \\ 
  & (0.262) & (0.189) & (0.215) \\ 
  & & & \\ 
\hline \\[-1.8ex] 
Observations & 6,090 & 7,063 & 1,896 \\ 
Log Likelihood & $-$2,333.702 & $-$2,769.613 & $-$943.108 \\ 
Akaike Inf. Crit. & 4,673.403 & 5,545.226 & 1,892.215 \\ 
\hline 
\hline \\[-1.8ex] 
\textit{Note:}  & \multicolumn{3}{r}{$^{*}$p$<$0.1; $^{**}$p$<$0.05; $^{***}$p$<$0.01} \\ 
\end{tabular} 
\end{table}

% Table created by stargazer v.5.2.2 by Marek Hlavac, Harvard University. E-mail: hlavac at fas.harvard.edu
% Date and time: Tue, Jan 01, 2019 - 13:51:50
\begin{table}[!h] 
\centering 
  \caption{Robustness Checks} 
  \label{} 
\begin{tabular}{@{\extracolsep{5pt}}lcc} 

 & \multicolumn{2}{c}{\textit{Dependent variable:}} \\ 
\cline{2-3} 
\\[-1.8ex] & \multicolumn{2}{c}{wiki\_bool} \\ 
\\[-1.8ex] & (1) & (2)\\ 
\hline \\[-1.8ex] 
 gendermale & 0.546$^{***}$ & 0.480$^{***}$ \\ 
  & (0.071) & (0.071) \\ 
  & & \\ 
 H5 index & 0.548$^{***}$ &  \\ 
  & (0.034) &  \\ 
  & & \\ 
 Citation Count &  & 0.325$^{***}$ \\ 
  &  & (0.016) \\ 
  & & \\ 
 Constant & $-$3.619$^{***}$ & $-$4.552$^{***}$ \\ 
  & (0.110) & (0.133) \\ 
  & & \\ 
\hline \\[-1.8ex] 
Observations & 15,049 & 15,049 \\ 
Log Likelihood & $-$6,423.468 & $-$6,335.015 \\ 
Akaike Inf. Crit. & 12,852.940 & 12,676.030 \\ 
\hline 
\hline \\[-1.8ex] 
\textit{Note:}  & \multicolumn{2}{r}{$^{*}$p$<$0.1; $^{**}$p$<$0.05; $^{***}$p$<$0.01} \\ 
\end{tabular} 
\end{table}

% Table created by stargazer v.5.2.2 by Marek Hlavac, Harvard University. E-mail: hlavac at fas.harvard.edu
% Date and time: Tue, Jan 01, 2019 - 12:49:07
\begin{table}[!htbp] \centering 
  \caption{Summary Statistics for Continuous Variables } 
  \label{tab:sumstats} 
\begin{tabular}{@{\extracolsep{5pt}}lccccccc} 

Statistic & \multicolumn{1}{c}{N} & \multicolumn{1}{c}{Mean} & \multicolumn{1}{c}{St. Dev.} & \multicolumn{1}{c}{Min} & \multicolumn{1}{c}{Pctl(25)} & \multicolumn{1}{c}{Pctl(75)} & \multicolumn{1}{c}{Max} \\ 
\hline \\[-1.8ex] 
h.index & 16,098 & 23.224 & 20.606 & 1 & 11 & 29 & 258 \\ 
h5.index & 16,098 & 16.943 & 14.965 & 0 & 8 & 21 & 191 \\ 
n.citations & 16,098 & 5,090.149 & 15,786.540 & 1 & 537 & 3,949.8 & 911,692 \\ 
\hline \\[-1.8ex] 
\end{tabular} 
\end{table}

\begin{figure}%[tbhp]
\centering
\includegraphics[width=0.8\linewidth]{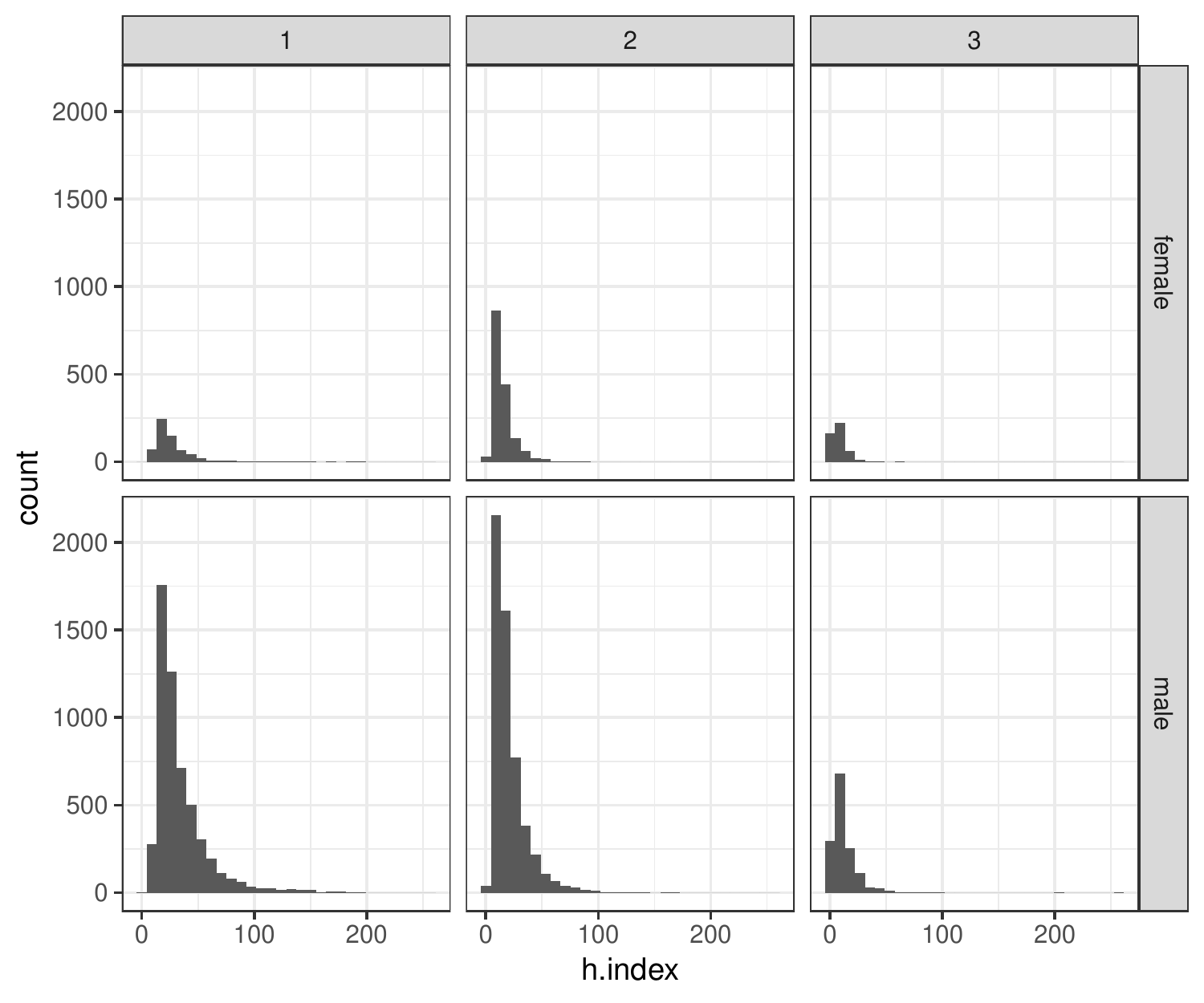}
\caption{Distributions of h-index by gender and field. 1 = Physics, 2 = Economics, 3 = Philosophy.}\label{fig:s1}
\end{figure}

\begin{figure}%[tbhp]
\centering
\includegraphics[width=0.8\linewidth]{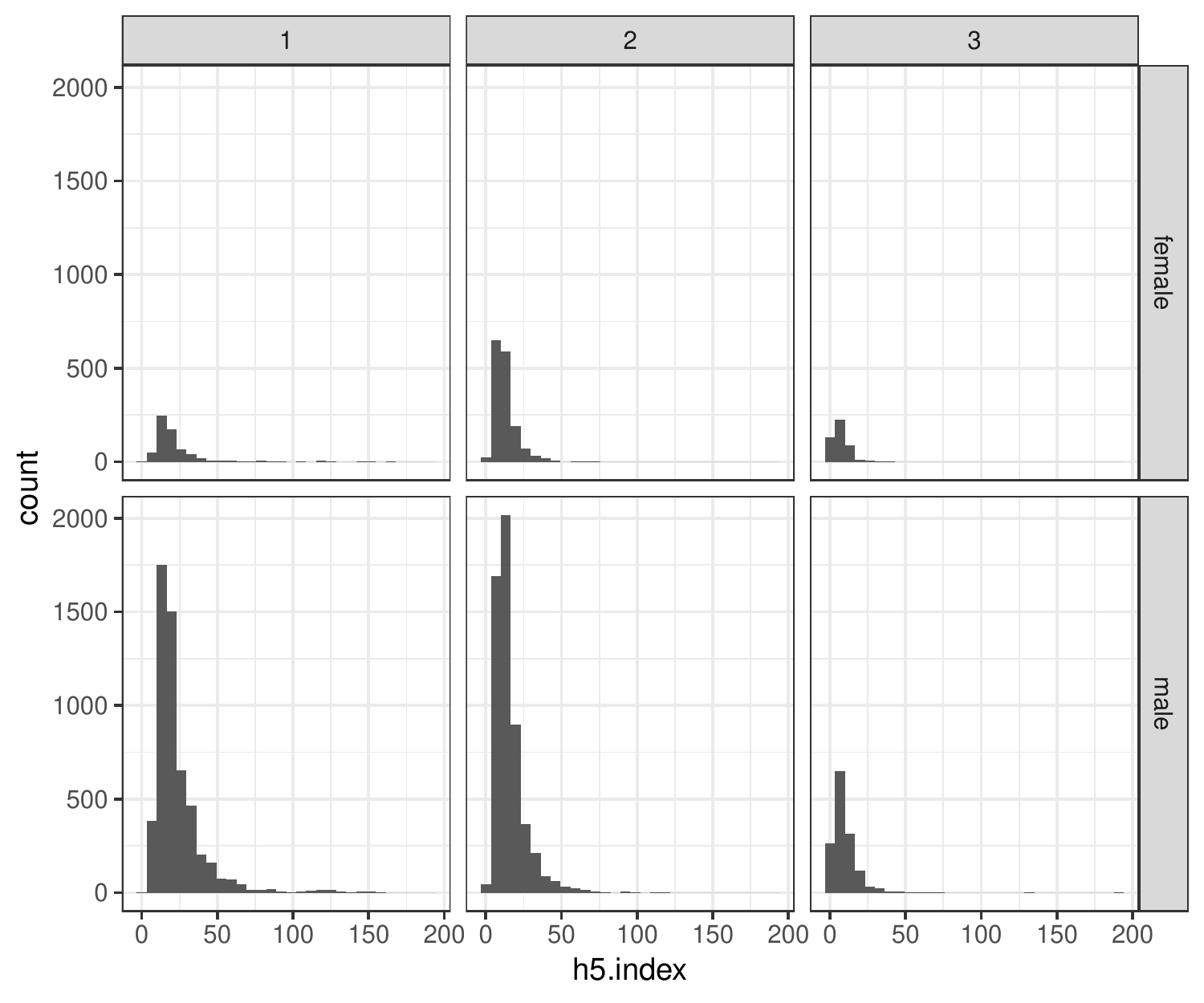}
\caption{Distributions of h5-index by gender and field. 1 = Physics, 2 = Economics, 3 = Philosophy.}\label{fig:s2}
\end{figure}

\begin{figure}%[tbhp]
\centering
\includegraphics[width=0.8\linewidth]{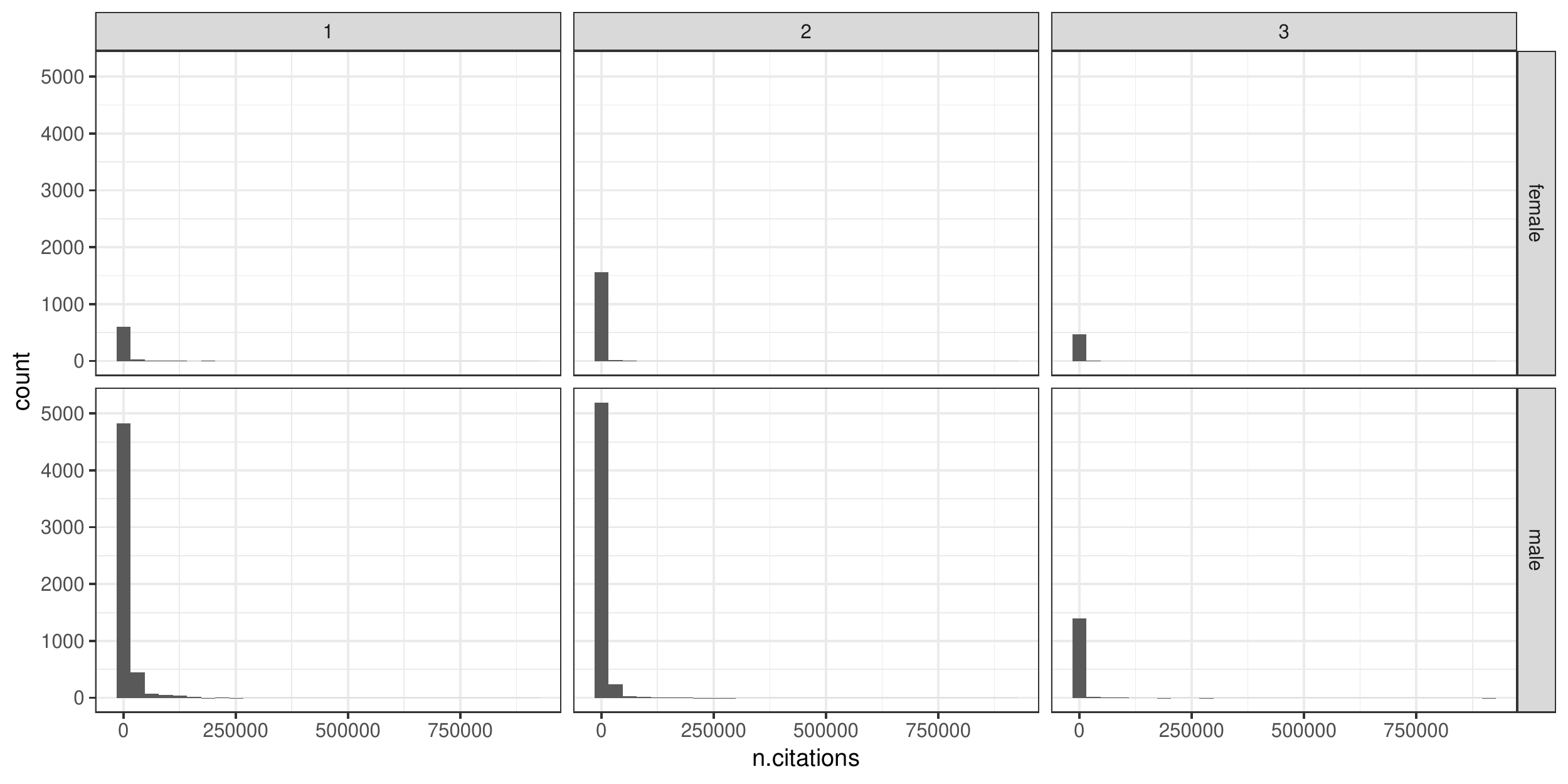}
\caption{Distributions of citations by gender and field. 1 = Physics, 2 = Economics, 3 = Philosophy.}\label{fig:s3}
\end{figure}

\end{document}